\documentclass[a4paper,preprint] {revtex4-1}
\pdfoutput=1

\usepackage{amssymb,amsmath}
\usepackage[T1]{fontenc}
\usepackage[utf8]{inputenc}
\usepackage{graphicx}   
\usepackage[english]{babel}

\newcommand{\beq}{\begin{equation}}
\newcommand{\eeq}{\end{equation}}

\begin{document}

\title{Power-law distribution in the number of confirmed COVID-19 cases}

\author{Bernd Blasius}
\email{blasius@icbm.de}
\homepage{http://staff.uol.de/bernd.blasius}

\affiliation{
Institute for Chemistry and Biology of the Marine Environment, University of Oldenburg, Oldenburg, Germany\\
ORCID: 0000-0002-6558-1462\\
}

\affiliation{Helmholtz Institute for Functional Marine Biodiversity (HIFMB), Carl von Ossietzky University Oldenburg, Oldenburg, Germany}

\date{\today}

\begin{abstract}   
\noindent

COVID-19 is an emerging respiratory infectious disease caused by the coronavirus SARS-CoV-2. 
It was first reported on in early December 2019 in Wuhan, China and 
within three month spread as a pandemic around the whole globe.
Here, we study macro-epidemiological patterns along the time course of the pandemic.
We compute the distribution of confirmed COVID-19 cases and deaths for countries worldwide and for
counties in the US, and show that both distributions
follow a truncated power-law over five orders of magnitude.
We are able to explain the origin of this scaling behavior as a dual-scale process:
the large-scale spread of the virus  between countries 
and the small-scale accumulation of case numbers within each country.
Assuming exponential growth on both scales,
the critical exponent of the power-law is determined by the ratio of large-scale to small-scale growth rates.
We confirm this theory in numerical simulations in a simple meta-population model, 
describing the epidemic spread in a network of interconnected countries. 
Our theory gives a mechanistic explanation why 
most COVID-19 cases occurred within a few epicenters, at least in the initial phase of the outbreak.
Assessing how well a simple dual-scale model predicts the early spread of epidemics, 
despite the huge contrasts between countries,
could help identify critical temporal and spatial scales of response in which to mitigate future epidemic threats. 


\end{abstract}

\maketitle

\section*{Introduction}

COVID-19 is an emerging infectious disease caused by the coronavirus SARS-CoV-2. 
It was first reported on in Hubei, mainland China on 31 December 2019 and has spread
well outside China in a matter of a few weeks, reaching 
countries in all parts of the globe within a time span of three month. 
As of 29 March 2020, the disease has arrived in 177 countries,  
with more than 700,000 confirmed cases and 30,000 deaths worldwide \cite{WHO}.
Despite the drastic, large-scale containment measures implemented in most countries 
these numbers are rapidly growing every day -
posing an unprecedented threat to the global health and economy of interconnected human societies. 

One of the most powerful tools to understand the laws of epidemic growth is mathematical modeling, going back
to Daniel Bernoulli's work \cite{Bernoulli_1760}
on the spread of smallpox in 1760.
Epidemiological models can be roughly divided into two classes.
The first class of models is focused on describing the 
temporal development of the epidemic within a localized region or country.
These models are often variants of the well known SIR-model
\cite{Kermack_1927, Keeling_Rohani_2008} and have recently been adapted to
the situation of COVID-19, taking into account non-pharmaceutical interventions (e.g., quarantine, hospitalization, and containment policies) and
allowing first predictions of healthcare demand  \cite{Ferguson_2020, MaierBrockmann_2020, Wang_2020}.

The second class of models is concerned with the geographic spread of the epidemic around the globe.
For these aims 
spatially explicit models have been developed that leverage information on the topology of transport networks. 
For example, the global network of cargo ship movements \cite{Kaluza_2010} was used to model the dispersal of invasive species \cite{Seebens_2013}. 
Similarly for infectious diseases, in a pioneering study, the 2003 spread of SARS in the global aviation network \cite{WooleyMeza_2011} was modeled \cite{Hufnagel_2004}.
Based on these approaches, conceptual frameworks have been developed to estimate epidemic arrival times as effective distances \cite{Brockmann_2013}.
At the same time, these models have been refined to highly detailed 
simulation frameworks for predicting the spread of disease
and are able to include factors such as vaccination,  multiple susceptibility classes, 
seasonal forcing, and the stochastic movement of individual agents
\cite{Colizza_2006, Broeck_2011}.
Reacting rapidly to the emergent pandemic, spatial epidemiological models
have been developed to describe and anticipate the spread of COVID-19
\cite{Arenas_2020, Chinazzi_2020, Gilbert_2020, Pullano_2020}.
These models allow to predict the incidence of the epidemics in a spatial population through time, permitting to study
the impact of travel restrictions and other control measures. 

Despite this theoretical progress, not much is known about the biogeography of COVID-19,
neither from empirical studies nor from mathematical models.
This is astonishing, as  one prominent characteristic 
of the pandemic 
is the huge variation in the number of cases that have been reported from different countries of the world. 
As of March 2020, some few countries - the epicenters of the pandemic -
were already badly affected by the pandemic, while others at the same time had just confirmed the first few cases.
This geographic variation in COVID-19 prevalence might be explained by several arguments:
A first obvious possibility would be that the variation
is caused by the idiosyncratic circumstances of the individual countries
which differ largely in
their geography and population size, but also in the way they are combatting the disease.
Alternatively, parts of the variation could simply be due to reporting errors,
reflecting disparate national testing regimes,
with countries such as  China, Japan, South Korea, or Germany having high testing rates, in contrast to other countries with 
much poorer testing. 
Here, we argue, however, that a dominant part of this variation
may be a direct consequence of the dynamics of the spreading process itself.
Thereby, the epidemic prevalence in a country should be directly correlated to the arrival time of the disease:
countries that were invaded very early by the virus 
have accumulated many cases in time, while countries 
with a late invasion naturally still have smaller prevalence.

To test this hypothesis, we use empirical data \cite{Dong_2020}
to compute the country-level distribution, $P$, of confirmed COVID-19 cases, $n$, on end of March 2020 worldwide
and find that it is closely approximated by a truncated power-law
\begin{equation}
P(n) \sim n^{-\mu}, \quad 1 \leq n \leq n_{max}
\label{Eq:PL}
\end{equation}
over five orders of magnitude.

Power-law distributions characterize a large range of phenomena in natural, economic, and social systems,
which is known as Zipf- or Pareto law
 \cite{Sornette_2003, Mitzenmacher_2004, Newman_2005, Clauset_2009}.
Examples range from the number of species in biological taxa \cite{Yule_1925}, 
the number of cities with a given size \cite{Zipf_1949}, the number of different words in human language \cite{Zipf_1949},
the frequency of earthquakes \cite{Gutenberg_1944},
the distribution of wealth \cite{Pareto_1964}, 
the number of scientific citations \cite{Price_1965, Redner_1998},  
the step length in animal search patterns \cite{Viswanathan_1996}, 
and the popularity of chess openings \cite{Blasius_2009}.
Our study shows that epidemic prevalence, at least in the emerging stage of a pandemic, is another 
system that falls into this class,
suggesting that the spatial distribution of COVID-19 case numbers is a fractal \cite{Bunde_2012}.

The appearance of a power-law distribution often points to 
the nature of the underlying processes.
It might, for example, be an indication that the system operates close to criticality \cite{Bak_1987, Newman_2005},
it might hint at the presence of
a multiplicative stochastic process with certain boundaries \cite{Sornette_2003, Blasius_2009},
or a rich-get-richer process \cite{Simon_1955, Barabasi_1999}.
Here, we provide a conceptual dual-scale model that explains the emergence of the power-law distribution 
by the 'superposition' of two concurrent processes:
large-scale spread of the virus  
between countries and small-scale snowballing of case numbers within each country.
Assuming exponential growth on both scales,
the critical exponent is simply determined by the ratio of large-scale to small-scale growth rates.
We confirm this theory in
numerical simulations in a simple meta-population model, describing the epidemic spread in a 
network of interconnected countries. 
By combining real world data, modeling, and numerical simulations we make the case that that the distribution of epidemic prevalence,
and possibly that of spreading processes in general, might follow universal rules.

\section*{Results}

\begin{figure}[t]
\center \includegraphics[width=0.9\textwidth]{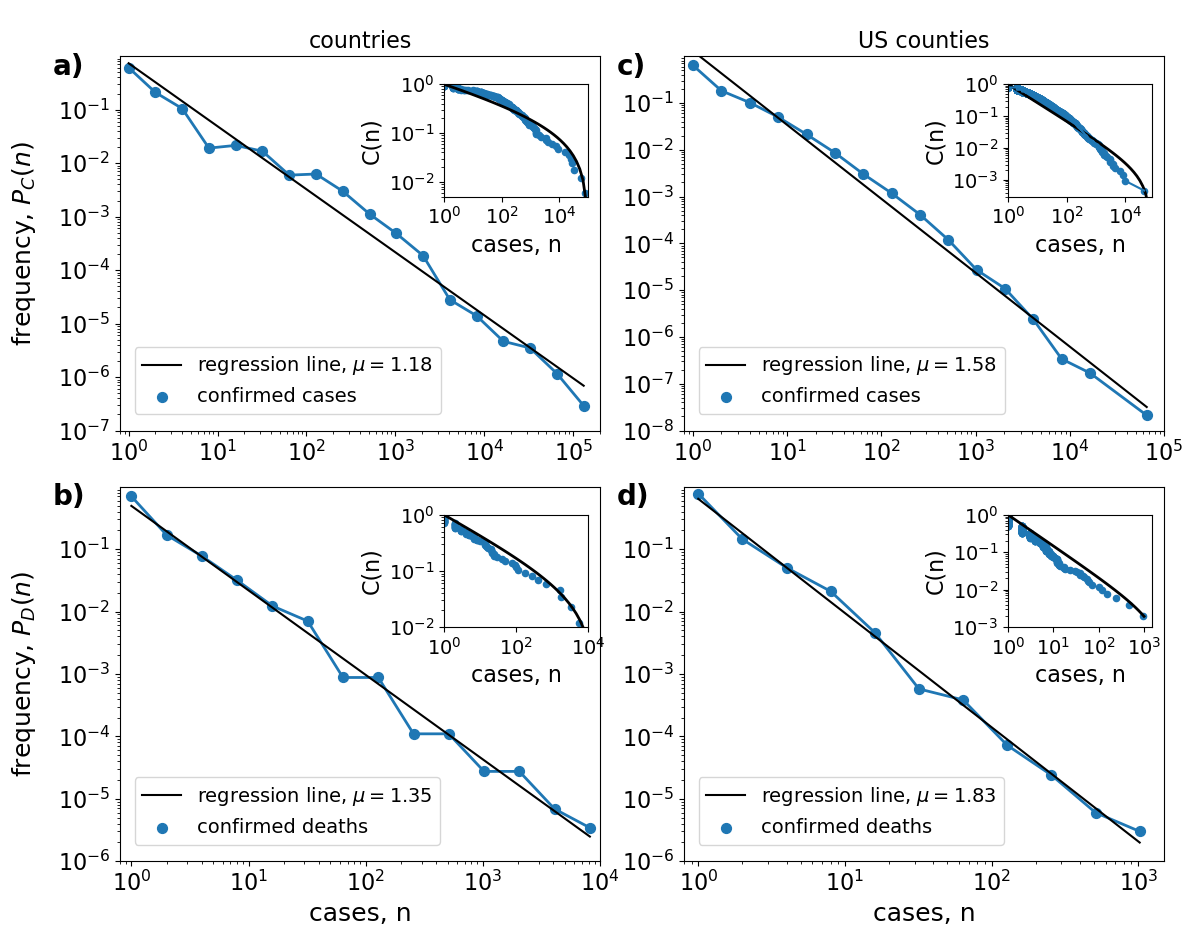}
\caption{Power-law scaling in the distribution of confirmed COVID-19 cases.
Left column: Estimated probability $P_x(n)$ (blue lines and circles) for a country to have a certain number $n$ of 
{\bf (a)} confirmed cases ($x=C$) and
{\bf (b)} confirmed deaths ($x=D$) on 22 March, 2020.
Right column: the same for the 2160 US counties that have been invaded by the coronavirus on March 31, 2020.
Histogram bins are spaced equally on a logarithmic axis
and only bins with a positive number of entries are shown.
Black solid lines show straight-line fits with slope $\mu$, indicated in the figure labels.
Insets: Cumulative fraction $C(n) =\sum_{m=n+1}^N  P(m) $ 
of countries, or counties, with case number $m > n$.
Solid lines show the cumulative distribution Eq.~(\ref{Eq:Cumul})
of a truncated power-law distribution with critical exponent $\mu$ and cut-off value
{\bf (a)} $n_{max}=1 \cdot 10^5$,
{\bf (b)} $n_{max}=1.5 \cdot 10^4$, 
{\bf (c)} $n_{max}=7 \cdot 10^4$, and
{\bf (d)} $n_{max}=3 \cdot 10^3$.
}
\label{fig:distributions_all}
\end{figure}

\subsection*{Power-law distribution in empirical data} 

Our research builds on the COVID-19 data repository
operated by the Johns Hopkins University Center for Systems Science and Engineering (JHU CSSE) \cite{Dong_2020}.
The database contains information about the daily number of confirmed  COVID-19 cases and confirmed deaths 
in various countries worldwide.
Using this data we computed the distribution $P_C(n)$ of confirmed cases and the distribution $P_D(n)$ of confirmed deaths
at a given date (see Methods).
The country-level prevalence distribution on March 22, 2020 
is shown in Fig.~\ref{fig:distributions_all} a, b.
On that day 168
countries  were invaded by the coronavirus and
86 countries already had reported fatalities.
The number of confirmed cases varied between 81,435 cases in China 
(followed by 59,138 cases in Italy) and 1 case in 16 countries.
The number of confirmed deaths varied between 5,476 in Italy (followed by 3,274 in China)
and one or zero deaths in many countries.
Figs.~\ref{fig:distributions_all}a,b clearly demonstrate that the frequency $P$ of countries that have a certain number $n$ of COVID-19 cases
follows a broad, long-tailed distribution that in very good approximation can be described by a 
power-law, spanning five orders of magnitude for the confirmed number of cases and
four orders of magnitude for the confirmed number of deaths.

To illustrate the robustness of our hypothesis to spatial scale, 
in Fig.~\ref{fig:distributions_all}c, d we depict the same analysis for the distribution
of confirmed COVID-19 cases in US counties on March 31, 2020.
On this day 2160 counties were invaded by the virus and 514 counties reported at least on death. 
Epidemic prevalence
varied between 43,119 confirmed cases and 922 confirmed deaths in New York City and
one confirmed case in 455 counties and one confirmed death in 253 counties.
Again, we find that the distribution of confirmed cases follows a power-law over several orders of magnitude.
Thus, although the two data sets differ greatly in spatial scale and resolution
(168 invaded countries in Fig.~\ref{fig:distributions_all}a,b
vs. 2160 invaded US counties in Fig.~\ref{fig:distributions_all}c,d)
we obtain very similar patterns of prevalence distribution.

A crude estimation of the critical exponent can be obtained by measuring
the slope of a regression line through the data on a double-logarithmic plot.
Applying this method to the country-level distribution (Fig.~\ref{fig:distributions_all}a,b),
we obtain a value of
$\mu_C = 1.18$ (slope of the distribution of confirmed cases) and
$\mu_D = 1.35$ (confirmed deaths). 
For the US-county distribution (Fig.~\ref{fig:distributions_all}c,d) we obtain somewhat larger slopes 
of $\mu_C = 1.58$ and $\mu_D = 1.83$.
A more accurate estimation of the critical exponent 
is provided by a maximum likelihood estimation (see Methods).
Applying this approach to the country-level COVID-19 distribution 
yields  critical exponents of
$\hat{\mu}_C=1.14 \pm 0.01$
and $\hat{\mu}_D=1.50 \pm 0.05 $.
For the US-county distribution we obtain the value $\hat{\mu}_C = 1.49 \pm 0.01$
and $\hat{\mu}_D=2.31 \pm 0.06$.
These exponents slightly deviate from 
those obtained from the regression analysis, 
but are still in the same ballpark.

Given an unbounded 
power-law distribution $P(n)$, 
the cumulative distribution function
$C(n) = \int_n^\infty P(n') dn'$ should also follow a power-law 
$C(n) \sim n^{1-\mu}$.
As shown in the insets in Fig.~\ref{fig:distributions_all}, 
this is not the case for the distribution of COVID-19 cases,
for which
the cumulative fraction $C(n) =\sum_{m=n+1}^N  P(m) $ 
of countries, or counties, with case number $m > n$
do not really follow a straight line in a double logarithmic plot.
Instead, they are better described by the cumulative distribution function
Eq.~(\ref{Eq:Cumul}) of a truncated power-law,
that is, a power-law distribution with an upper bound $n_{max}$ for the number of cases,
Eq.~(\ref{Eq:PL})
(see Methods and Appendix Fig.~\ref{fig:robustness}).
This indication for the presence of a truncated power-law distribution is also conform with our theoretical analysis below.

However we note that although the shape of the empirically obtained $C(n)$ in overall follows the curve of a truncated power-law distribution,
there is a considerable wavering around the theoretical curve 
(compare blue circles and black lines in Fig.~\ref{fig:distributions_all} insets).
Thus, a rigorous hypothesis testing with Monte Carlo simulations
\cite{Clauset_2009, Deluca_2013},
which does not take disturbances due to additional irregularities (e.g., heterogeneities in country sizes or containment measures)
fully into account,
will always reject the hypothesis of a perfect truncated power-law as the true underlying distribution.

The presence of a power-law distribution means that global COVID-19 prevalence patterns are 
characterized by a small number of countries with huge epidemic prevalence (the long tail of the distribution) 
and a large number of countries that are (yet) barely affected by the disease.
In between these two extremes there is a smooth transition and this transition is
scale free, that is, the amplification in the number of countries (or counties) with decreasing number of cases
is the same at all scales.
In general, the obtained critical exponents are rather small.
While  for most natural power-law distributions critical exponents are around $\mu \approx 2$,
here we estimate exponents that are clearly below two,
$\mu<2$, indicating a very broad distribution for which in the absence of an upper bound the mean value diverges.

\begin{figure}[th]
\center \includegraphics[width=0.7\textwidth]{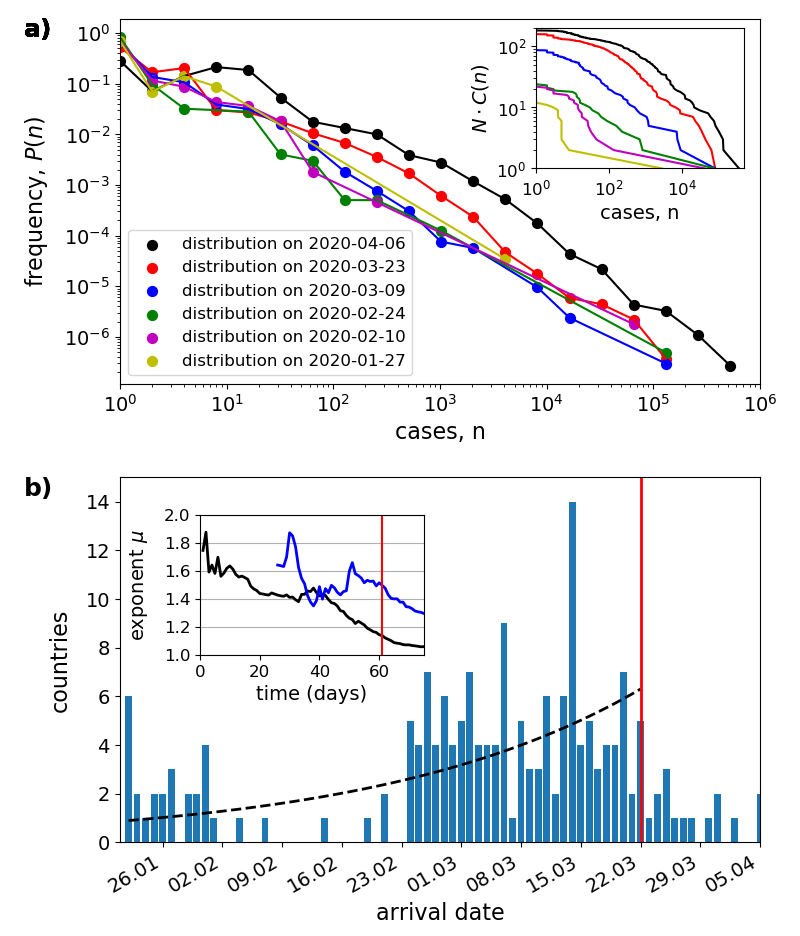}
\caption{
Temporal development of the COVID-19 pandemic.
{\bf a)} Evolution of the distribution of confirmed  cases per country.
The same as Fig.1a, but for six different time instances separated by 2 weeks (see figure legend) 
during the pandemic.
The inset shows the cumulative number of countries $N \cdot C(n)$, where
$N$ is the total number of countries with confirmed cases at that date.
{\bf b)} Distribution of arrival times.
The histogram shows the number of countries that were invaded by the virus
on a certain day between Jan 22, 2020 and April 5, 2020 (blue bars). 
Further shown is an exponential increasing function, $\exp(s t)$ (black dashed line) with growth rate
$s=0.03 \ d^{-1}$, obtained by a least square fit to the histogram during the first 61 days.
The inset shows the critical exponents $\hat{\mu}_C(t)$ (black) and $\hat{\mu}_D(t)$ 
(blue, only shown from 16 February, the first days with at least 5 confirmed deaths), estimated by
maximizing the log-likelihood function
Eq.~(\ref{Eq:LogLikeFun}), as a function of time.
The vertical red line indicates 22 March, the date of the distribution shown in Figs.~\ref{fig:distributions_all}a,b.
}
\label{fig:buildup}
\end{figure}

\subsection*{Temporal development during the pandemic spread}

While the present analysis considers the distribution of case numbers at a temporal snapshot, in reality
the pandemic is a dynamic process successively invading countries worldwide.
In Fig.~\ref{fig:buildup} we investigate the temporal development of the COVID-19 distribution.
The figure shows that the country-level distribution of confirmed cases is formed
already within a few weeks from the start of the outbreak 
and remains roughly stationary over the considered time interval of 75 days.
A closer inspection
(see inset in Fig.~\ref{fig:buildup}b) reveals that the critical exponents in fact are not constant, 
but in general are decreasing functions of time,
indicating that the case distributions tend to broaden over the course of the pandemic.


Fig. \ref{fig:buildup}b further investigates the spatial spread of COVID-19 
across countries worldwide more systematically.
The figure plots the number of countries that were invaded by the coronavirus 
(i.e., having the first confirmed COVID-19 case) 
at a particular day  in the time span from January 22 to April 5, 2020.
On January 22, the first entry in the database, six countries 
(China, Japan, South Korea, Taiwan, Thailand, US) were already invaded by the virus.
From this day, within roughly two months the pandemic spread to nearly every country in the world.

Interestingly, the invasion speed was not constant. Instead Fig. \ref{fig:buildup}b clearly
indicates two broad modes in the arrival time distribution.
A first group of countries was invaded by the disease in the end of January. 
In the first three weeks of February nearly no new arrivals were reported.
Starting from February 24, a second wave of invasions appeared
which lasted until end of March, after which the number of new arrivals began to fall again, 
probably reflecting the fact that 
the pandemic had reached basically all countries of the world.
As of April 5,  a total 185 countries were invaded by the coronavirus.

There are several possible reasons why the disease arrival is not more evenly distributed.
One explanation for the bimodal shape 
is related to the lockdown of airline transportation in China in the end of January 2020.
According to this hypothesis, after the first pandemic bubble in January, 
the further spread of the pandemic came to a temporary standstill with the onset of travel restrictions, 
only to resurface in a second wave, starting end of February.
Alternatively, it may be that many arrivals of the virus 
in countries all over the world simply went undetected
during the first weeks of February and
were detected only later with the increasing awareness and increased testing.
This hypothesis is corroborated by the observation that
end of February is also the time when the first PCR based tests became available.
In general, the strong irregularity in the arrival time distribution points to the 
high level of stochasticity of the worldwide spreading process.

\subsection*{Mechanistic explanation of the power-law distribution}

Fig.~\ref{fig:buildup} would suggest that the temporal development of the pandemic
is characterized by two complementary processes:
the successive invasion of more and more countries
and the increasing number of cases within each affected country.
Here we argue that the emergence of the power-law distribution
could be related to the concurrent 'superposition' of these two processes.
Thereby, on a large geographic scale, the pandemic is driven by the spread of the virus
in the network of interconnected countries.
On a small scale, case numbers are snowballing within each country, once it has been invaded,
thereby further increasing the epidemic imbalance due to different arrival times between countries. 

In the simplest approximation, at the begin of the pandemic
both of these processes developed exponentially in time. 
A straightforward calculation shows that the combination of the two exponential processes 
generically yields a truncated power-law distribution in the number of cases in countries:
Consider an epidemic outbreak that started (the first case reported in a country) at time $t=0$.
We are interested in the case distribution at time $t>0$.
Let us first assume that at this day
the probability for a country to have been invaded by the virus at some former time $\tau$
grows exponential in $\tau$ with spreading rate $s$,
\beq
P(\tau) \sim e^{s \tau}, \quad 0 \leq   \tau \leq t.
\label{Eq:ExpDistr}
\eeq
This exponential growth in the geographic distribution of the pandemic
would be the expectation if one modeled the spread in a network where nodes are countries
(neglecting saturation when the pandemic has reached most countries).
Note, that the
distribution is truncated from two sides because arrivals of the disease can only have occurred
after the start of the pandemic, $\tau \geq 0$, 
and in the past, $\tau \leq t$.

Second, we assume that in each country
the number of confirmed cases has grown exponentially with the time since invasion $t - \tau$
with growth rate $r$ (neglecting containment measures and saturation after the epidemic peak)
\beq
n(t) \sim e^{r(t-\tau)}.
\label{Eq:Exp}
\eeq
Combining these two equations, the probability distribution of confirmed cases $P(n)$ can be calculated as 
(see \cite{Newman_2005})
\beq
P(n) = P(\tau) \left|\frac{d \tau}{dn}\right|  \sim  \frac{e^{s \tau}}{e^{-r \tau}} \sim n^{-(1+s/r)}
, \quad \mathrm{with} \quad 1 \leq n \leq n_{max}
 \label{Eq:PowerLaw1}
\eeq
which is a truncated power-law with critical exponent 
\beq
 \mu = 1 + \frac{s}{r} .
 \label{Eq:law}
\eeq
Thus, the critical exponent is simply determined by the ratio of large-scale to small-scale growth rates.
In the symmetric case that both growth rates are identical, $s=r$, we would expect a power law with $\mu=2$.
In the limiting case that the large-scaling spreading process is linear in time, $s=0$, we obtain a border-line distribution with
critical exponent $\mu=1$.
Note, that from the truncation of $\tau$ in the arrival time distribution, Eq.~(\ref{Eq:ExpDistr}),
the admissible range of case numbers in the power-law distribution Eq.~(\ref{Eq:PowerLaw1})
necessarily is restricted between the lower bound $n=1$ 
(the epidemic prevalence in a newly invaded country)
and the cut-off value $n_{max} \sim e^{r t}$
(the epidemic prevalence at time $t$ in the country with the first confirmed case) -
justifying the observation of a truncated power-law in the empirical data as shown
in Fig.~\ref{fig:distributions_all}.

Obviously, this simple theory far from accurately describes a real-word pandemic. 
First of all, the theory is valid only in the initial phase of the pandemic, while both geographical spread and
within-country epidemic growth are still exponential. As soon as saturation processes set in, the derivation of the
power law breaks down.
Next, as shown in Fig.~\ref{fig:buildup}b the arrival time distribution during the COVID-19 pandemic is not exponential,
as discussed above.
In gross oversimplification we may nevertheless fit an exponential function
$P(t) \sim e^{st}$ through the data, yielding an 'average' spreading rate of $s=0.03 \ d^{-1}$ 
(black dashed line in Fig.~\ref{fig:buildup}b).
Finally, epidemic growth rates during the COVID-19 pandemic have not been not identical in all countries
(even in the initial stages). They have also not remained constant in time, but in most countries have fallen in 
the course of the epidemic.
Furthermore, most countries were invaded multiple times, leading to different epidemic foci within countries. 
Neglecting all these observations, for the sake of argument, let us 
assume an average doubling time of case numbers of 
$T_{1/2}= 3.5 \ d$ 
in all countries,
yielding an exponential growth rate of 
$r=\log(2)/T_{1/2}=0.2 \ d^{-1}$
and a maximal case number of $n_{max} = e^{0.2*60} = 1.6 \cdot 10^5$ after 60 days.
Then, according to our simple theory Eq.~(\ref{Eq:law})
we would expect a critical exponent of 
$\mu = 1 + 0.03/0.2 \approx 1.15$, 
in rather good agreement to the fitted exponents in Fig.~\ref{fig:distributions_all}.

\begin{figure}[t]
\center \includegraphics[width=0.8\textwidth]{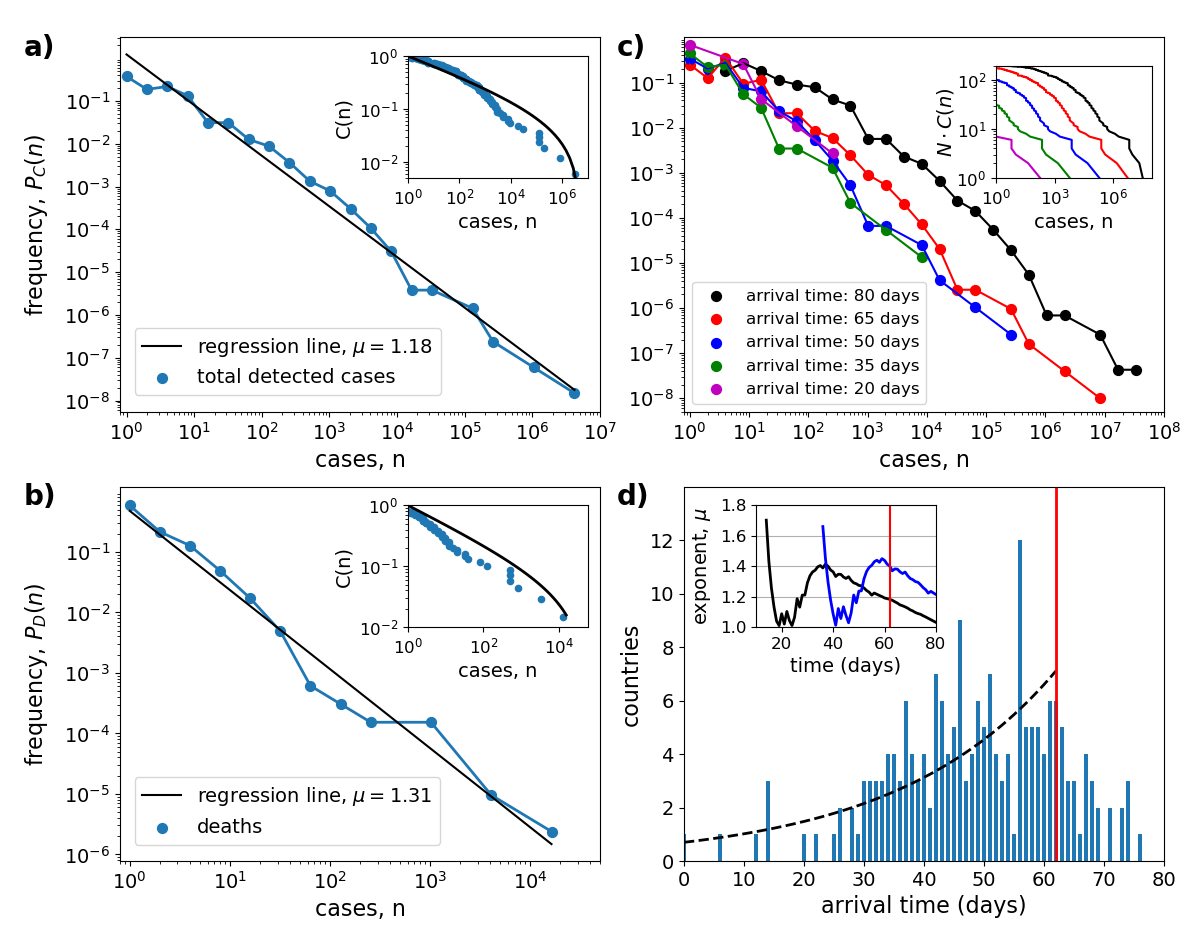}
\caption{
Power-law scaling in the meta-population model.
{\bf (a,b)} Same as Fig.~\ref{fig:distributions_all}, but for 
model simulations after transient of 62 days.
Shown is the estimated probability $P(n)$ (blue lines and circles) with straight-line fits (black lines) 
for the simulated cases {\bf (a)} and  deaths {\bf (b)}. 
Insets show the cumulative fraction $C(n)$ of countries (blue circles) and
the cumulative distribution function (black lines) of a truncated power-law distribution with
cut-off values  {\bf (a)} $n_{max}=5 \cdot 10^6$ and {\bf (b)} $n_{max}=5 \cdot 10^4$.
{\bf (c,d)} Same as Fig.~\ref{fig:buildup}, showing the spatial spread in the meta-population model.
{\bf (c)} Evolution of the distribution of cases.
as in {\bf (a)}, but for five different time instances separated by 15 days (see figure legend). 
{\bf (d)} Histogram of arrival times, showing the number of
countries that were invaded on a certain day. Simulation time: 80 days.
The black dashed line shows an exponentially increasing function, $\exp(s t)$ with spreading rate
$s=0.037 \ d^{-1}$, obtained by a least square fit to the data during the first 62 days.
The inset shows the time dependence of the critical exponents $\hat{\mu}_C(t)$ (black) and $\hat{\mu}_D(t)$ (blue)
for the distribution of the number of cases and deaths, estimated by maximizing the log-likelihood function
Eq.~(\ref{Eq:LogLikeFun}) for all days where at least 5 cases were reported.
The red vertical line indicates day 62, the time of the distribution in  {\bf (a,b)}.
See Methods for model description and parameter values.
}
\label{fig:meta}
\end{figure}

\subsection*{Results from a meta-population model}

To test the theory of the previous section, we developed a 
dual-scale meta-population model (see Methods).
The first level describes the large-scale stochastic spread of the virus in a network of $N$ interconnected countries.
The second level describes the small-scale increase in case numbers within a country;
it is started in each country from the time point of invasion by the virus 
and the follows a simple deterministic SIR-dynamics (see Methods).
The motivation for this model design was not to predict the worldwide spread of COVID-19, but rather
to quantitatively test the emergence of heterogeneous case distributions in a conceptual model framework that incorporates the
ideas from the previous section.

Fig.~\ref{fig:meta} shows a typical model outcome.
The large-scale spreading process is captured in the arrival time distribution
which exhibits a unimodal dependency on time  (Fig.~\ref{fig:meta}d).
Correspondingly, the number of invaded countries grows stochastically and roughly follows a sigmoidal shape.
In accord to our theory, Eq.~(\ref{Eq:ExpDistr}),
this arrival time distribution starts to grow exponentially in the build-up phase of the pandemic.
The highest invasion rates occur after about 50 days, while 
after a simulation time of 80 days 196 out of the $N=200$ countries are already invaded by the virus.

Combining the large-scale and small-scale model components allows to simulate the epidemic prevalence
in each country as a function of time.
Fig.~\ref{fig:meta}a,b shows the resulting distribution of cases and deaths after a simulation time of 62 days 
(vertical red line in Fig.~\ref{fig:meta}d).
Again, the distributions are well characterized by a truncated power-law.
Comparison with Fig.~\ref{fig:distributions_all} shows that the
model is able to describe the characteristics of the empirical distribution of COVID-19 cases rather well.
The log-likelihood estimation of the critical exponents yields 
values of $\hat{\mu}_C=1.18 \pm 0.01$ and $\hat{\mu}_D=1.40 \pm 0.06$.
These exponents can be compared to our theory Eq.~(\ref{Eq:law}).
From Fig.~\ref{fig:meta}d we estimate a spatial spreading rate of 
$s=0.037 \ d^{-1}$ in the build-up phase of the pandemic.
The initial growth rate of infected in the SIR-model equals $r=0.23 \ d^{-1}$ (see Methods).
Thus,  according to  Eq.~(\ref{Eq:law})
we would expect a critical exponent of
$\mu = 1 + 0.037/0.23 = 0.16$, in good agreement to the estimated value from the numerical
simulation.

We want to note that the nearly ideal power-law scaling in the case distribution holds only in the initial phase of the pandemic and
 is lost when the spatial spreading starts to saturate. 
This can be seen in the simulated case distribution $P(n)$ for different time instances
(Fig.~\ref{fig:meta}c).
While $P(n)$ remains roughly stationary for the first 50 to 60 days of the simulation, a first plateau begins to emerge
at the left end of the distribution for larger times. This plateau reflects the fact that when the number of newly invaded countries 
is reduced, these countries with just a few cases are missing in the left end of the case distribution
(reminiscent to the behavior exhibited in the empirical case distribution, Fig.~\ref{fig:buildup}a).
Additionally, the estimated critical exponents exponents are decaying in time (Fig.~\ref{fig:meta}d inset),
similar to that of the empirical data (Fig.~\ref{fig:buildup}b inset).
Thus, the first sign that the outbreak has reached most countries in the network
is the  reduction in the scaling range and a simultaneous broadening of the case distribution. 
Eventually, in the limit of large time, 
when the epidemic has come to an end in every country,
scaling is lost and
the distribution of cases must converge towards a delta function $P(n) = \delta(n-fN_{pop})$,
with $f$ the fraction of susceptible out of a population of $N_{pop}$ individuals in a country that will be infected
(or it would approach the country size distribution in a meta-population with heterogeneously distributed country sizes).
Interestingly, in our numerical simulations, we still obtained power-law distribution 
when the contact rate $\beta$ was set to a large value, so
that the dynamics within a country rapidly reach a stationary state. In this case, with increasing 
$\beta$ (and thus  increasing initial epidemic growth rates $r$)
the critical exponents tended to $\mu \to 1$.

\section*{Discussion}

It is well known from the literature (e.g., \cite{Newman_2005, Clauset_2009})
that caution is in order when trying to identify power-law distributions in real data and, in particular,
that a straight line in a double-logarithmic plot does not suffice to prove the existence of a power law distribution.
For this reason, the aim of this study is not to
prove that the COVID-19 case distribution is a perfect power-law,
an undertaking that would require sophisticated statistical analysis 
and a much larger sample size \cite{Clauset_2009}.
We also do not intend to rule-out
other likely candidate distributions (e.g., log-normal  or stretched exponential distributions).
Instead, our claim is merely to demonstrate that the empirical data are {\em highly consistent with the 
hypothesis} that the number of reported cases are taken from a truncated power-law distribution of the form Eq.~(\ref{Eq:PL}).

\begin{figure}[t]
\center \includegraphics[width=0.8\textwidth]{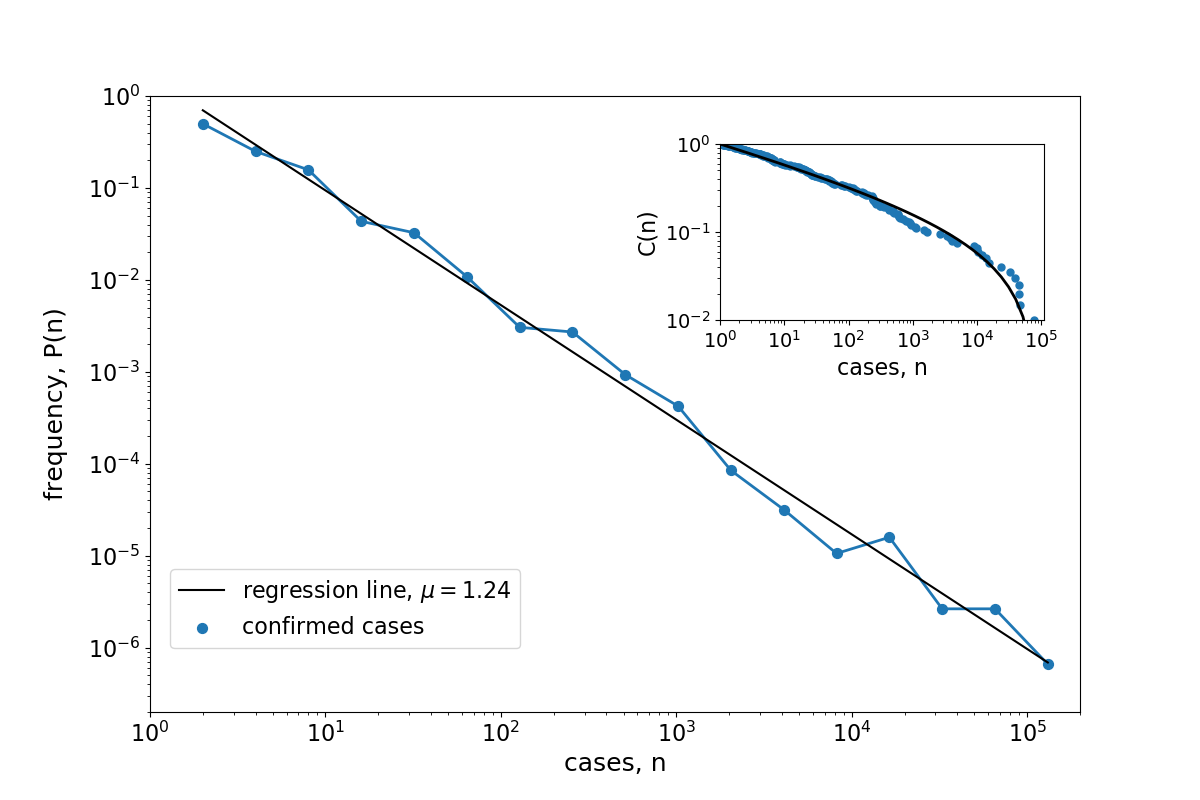}
\caption{
Robustness of the algorithm for estimating parameters of a truncated power-law.
The same as Fig.~\ref{fig:distributions_all} but for 200 random numbers $n_i$ that were generated from a truncated power-law distribution with 
$\mu = 1.2$ and cut-off value $n_{max}=1\cdot10^5$ (see Methods).
The estimated distribution roughly follows a straight line on the double-logarithmic plot with equally spaced bins. Note, that even though only 200 random numbers were drawn, the estimated probabilities vary over many orders of magnitude (which is numerically possible since in order to compute the probability distribution, the histogram counts are divided by the variable bin sizes). 
Log-likelihood estimation of critical exponent (Methods) yields $\hat{\mu}= 1.21 \pm 0.01$ in good agreement with the actually used exponent. 
In contrast, the estimator for an unbound power-law, Eq.~(\ref{Eq:Loglike}), yields $\hat{\mu}= 1.28 \pm 0.01$, strongly
overestimating the true exponent. 
The estimation by a regression line, yielding $\mu= 1.24$, also is slightly too large.
The inset shows that the cumulative fraction is well described by the cumulative distribution function $C(n)$ of a truncated power law
with critical exponent $\hat{\mu}$ and $n_{max}=8.4\cdot 10^4$, the maximal $n_i$ value of the sample.
}
\label{fig:robustness}
\end{figure}

Nevertheless, the scaling relations in the distributions shown in Figs.~\ref{fig:distributions_all} 
are remarkably constant over the whole range of case numbers, stretching several orders of magnitude
with no obvious signs of saturation 
for either the range of small or large case numbers.
One might argue that the bend in the cumulative distribution function is a sign that the growth in some countries 
(e.g., China, Korea) had already become sub-exponential. 
However, this is contradicted by the observation that a similar bend is also exhibited by the cumulative
distribution function obtained from the meta-population model 
(Fig.~\ref{fig:meta} and).
Thus, the most likely explanation is that the case distribution follows a truncated power-law
(see also Fig.~\ref{fig:robustness}),
suggesting the hypothesis that the spatial distribution of COVID-19 cases is a fractal \cite{Bunde_2012}.
This is further corroborated by our simple theory which provides a mechanistic explanation for why we would
a expect a truncated power-law in the first place.

Our finding of power-law distributions in the number of reported cases has important consequences
for epidemiology.
Most notably, the small values of the estimated critical power-law exponents are
related to the strong inequality of case numbers that was frequently observed all over the world in the initial phase of
the COVID-19 outbreak.
Following a power law distribution means that this pattern prevails even as numbers grew and the scale of infection expanded globally.
In particular, during the course of the pandemic, most cases were reported to have occurred in a few countries,
sometimes even a single country - the so-called epicenters of the pandemic.
The distribution of cases within countries followed a similar pattern.
Often COVID-19 was peaking in a few localized foci (local regions or cities), 
while other parts of the country at the same time had experienced only a moderate number of cases.
Our theory provides a mechanistic explanation why this might have been the case.

A graphical representation for the inequality of a distribution is given by the Lorenz curve \cite{Newman_2005} which 
in the case of the COVID-19 case distribution is a plot of 
the fraction of the total number of confirmed cases 
in dependence of the fraction of the most affected countries.
This is shown in the Appendix Fig.~\ref{fig:lorenz} for the number of confirmed COVID-19 cases and confirmed deaths
on 22 March 2020.
The Lorenz curve shows that on this day 95.7\% of confirmed cases and  97.6\% of the confirmed deaths 
had been reported in the 20\% most affected countries 
(while the top 5\% most affected countries had accumulated 82.3\% of all confirmed cases and 84.4\% of all confirmed deaths).
With 81,435 out of 336,953 confirmed cases on that day China alone had accumulated a fraction of  24\% of all cases.
The two most affected countries, China and Italy, together had accumulated a fraction of 41\% of the worldwide reported cases.

This inequality can also be measured by the Gini-coefficient $G$ \cite{Gini_1912},
which ranges between $G=0$ for perfect equality, i.e., all countries having the same number of cases, 
and $G=1$, corresponding to maximal inequality, where all cases appear in a single country.
For the distribution of confirmed COVID-19 cases on March 22 (Fig.~\ref{fig:distributions_all}a,b)
we obtain a Gini-coefficient of $G=0.92$ and for the number of confirmed death of $G=0.94$.
These large values 
are a direct consequence of the  small critical exponents of the estimated power-law distributions. 
In fact, for an unbounded power-law distribution with
$\mu<2$ one would theoretically expect a Gini-coefficient of $G=1$ \cite{Newman_2005}.

The emergence of power-law distributions with a small critical exponent and the associated inequality of the distribution, 
with Gini coefficients close to one is also observed in the developed meta-population model.
Consequently, also in the model case numbers are mostly concentrated in a few countries.
In the simulations, these epicenters of the pandemic, i.e., the countries with most cases, are
always the countries in which the diseases originated or which were first invaded by the virus.
In other words, the prevalence rank order among countries remains unchanged during the course of the pandemic.
This is akin to the "rich-get-richer process" or "first-mover-advantage" \cite{Simon_1955, Newman2009},
a well-studied process to generate power-law distributions.
In the real COVID-19 pandemic, this was not the case. During the begin of the pandemic most cases were observed in China,
later the "leading role" changed next to Italy and finally to the USA. This reflects different mitigation strategies and circumstances in different countries, a factor that is not considered in the simple model.
Nevertheless, despite these changes in the rank order, the 
distribution of cases in the empirical data was always closely represented by a power-law.

We would like to remark that the available database only provides information on the reported COVID-19 cases
in each country.
In all likelihood, the real number of cases will be much larger. 
Not much is known about the reporting rates, but first estimates indicate that a substantial fraction (possible 86\%) of infections might go undetected \cite{Li_2020}. 
Reporting rates probably vary strongly between countries and
may change in time with the awareness of national health institutions and available testing capabilities. 
Further uncertainties arise because the criteria by which a person is classified as active case (and even more so for being 
classified as recovered) vary between countries and not uncommonly have been modified during the course of the pandemic within a country. 

Remarkably, we obtained power-law distributions in the absolute number of cases in each country.
At first guess, one might have expected such scaling only after case numbers have been normalized by
population sizes. Our preliminary investigations show that such normalized
case numbers become even more unequally distributed, with even smaller estimated values of the critical exponent,
 and the distributed values do not line-up any more so well on a straight line on a double logarithmic plot.
Thus 'folding'  the distribution of population sizes over 
the COVID-19 case distribution does not flatten, 
but rather tends to further increase, the inequality of the resulting distribution. 
This indicates that absolute (non-normalized) case numbers may be the
natural variables to describe the patterns of the pandemic in its initial stage.
In all likelihood, the role of country sizes and population numbers will become
increasingly  important with the further spread of the pandemic.



We have shown that a simple conceptual model yields an accurate description of the COVID-19 prevalence distribution
in the initial phase of the pandemic.
This is remarkable in that many important epidemiological aspects of the spreading process are not captured by the model.
Most notably,
the model does not 
neither take into account 
variability in country sizes, population numbers, or testing rates,
heterogeneity of intra- and inter-country 
connectivity, as well as the corresponding changes due to
social distancing, lock-down measures, closing of airline connections and shut-down of borders.
These simplifications leave much room for future investigations and model improvements.
Obvious model improvements would be to consider
a meta-population with  heterogeneously distributed country sizes or to
make the initial number
of infected individuals a random number, as would be a better description of what happened in many countries.


One basic assumption of the developed model is the separation of the pandemic into two spatial scales,
the large-spatial spread over a rather small number ($N<200$) of interconnected countries and
the small-scale growth within a population of much larger size ($N=5 \cdot 10^7$). 
This separation, obviously is somewhat arbitrary.
For the virus countries are, of course, quasi-arbitrary entities. 
Therefore, it would be important to check whether 
both the data analysis (Fig.~\ref{fig:distributions_all}) and the mathematical model are
robust to arbitrarily subdividing or lumping countries.
The very similar scaling observed among US counties (Fig.~\ref{fig:distributions_all}c,d)
lends credence to the model's generality.
Similarly, one can readily ascertain that the model result is not an artifact of artificial lumping. 
Suppose a virus that is spreading
in an all-to-all, or randomly coupled, network of a number of $N \cdot N_{pop}$ individuals.
If we would artificially subdivide individuals into a small number $N$ of classes (or countries), 
at the time point when the disease has spread to all countries, within each country we would still have only
a few cases (of the order of $N \ll N_{pop}$). Thus, the assumed simultaneous spread on both spatial scales requires
a real physical separation in the network structure.
It would be an interesting perspective for future research to study the spread in multi-scale hierarchies or in more
realistic models of interconnected societies.

One important model application would be the simulation of interim COVID-19 lockdown or containment measures, 
as were introduced in many countries in the world in March and April 2020. Such measures might inhibit the increase of case numbers within local regions (the small-scale part of our theory) but they would not necessarily suppress also the large-scale diffusion of infections across regions.
Thus, under the guise of suppressed case numbers during the mitigation period there could be a dangerous "invisible" homogenization in the spatial distribution of the virus. This would have tremendous implications in a scenario where the measures are suddenly lifted in many places. 
In this case our theory would predict the emergence of a very different case number distribution than shown in Fig.~\ref{fig:distributions_all}.
Instead of the previous power-law distribution resurfacing, 
the most likely situation would be 
the synchronous initiation of increasing in case numbers everywhere.
Thus, situations as they appeared only in the epicenters during the beginning phase of the pandemic 
could be the rule in most parts where mitigation measures are lifted.
In this sense, the long-tail of the case distribution, characterized by the many regions with only mild epidemic prevalence, that was observed in the initial phase of the pandemic, could create a false sense of security.

Finally, we would like to remark that the model's strong simplicity is at the same time a strength: 
being rather generic, it should be applicable to very different systems,
to describe the spread of commodities as a process with two spatial scales.
The fact that the distribution of COVID-19 resembles a model where only the initial infection 'counts' 
reflects the intrinsic difficulty in containing epidemics at global and local scales 
when unilateral measures (e.g., travel bans and lockdowns) are impractical or non-enforceable,  
i.e., where other countries or regions will step up and continue the spread. 
Thus, assessing how a well simple dual-scale model predicts the early spread of epidemics, 
despite the huge contrasts between countries, 
could help identify critical temporal and spatial scales of response in which to mitigate future epidemic threats. 


\section*{Methods}

\subsection*{Estimating parameters of a truncated power-law distribution}

Assume a truncated power-law (or Pareto) distribution of the random variable $n$
\beq
P(n) = C n ^{-\mu}, \quad 1 \leq n \leq n_{max}
\label{Eq:PLnorm}
\eeq
with upper bound $n_{max}$.
Normalization $\int P(n) dn=1$ yields $C=(1-\mu)/(n_{max}^{(1-\mu)}-1)$.\\
The cumulative distribution function reads
\beq
C(n) = \int_n^{n_ {max}} P(n') dn'  = \frac{ n^{(1-\mu)} - n_{max}^{(1-\mu)}} {1 - n_{max}^{(1-\mu)}}  .
\label{Eq:Cumul}
\eeq
In the limit $n_{max}\to \infty$ (a power-law distribution without upper bound)
the cumulative distribution function
also follow a power-law  $C(n) \sim n^{1-\mu}$.

A synthetic sample of the distribution (\ref{Eq:PLnorm}) can be obtained by the formula
\beq
n_i = \left[1 - u_i (1 - n_{max}^{(1-\mu)} ) \right]  ^{\frac{1}{1-\mu}} 
\label{Eq:SyntSamp}
\eeq
where $u_i$ are random numbers taken from a uniform distribution in the range $[0,1]$.
The inverse problem is to estimate the parameters of the distribution given a random sample 
$n(_1, n_2, \dots, n_N)$ of $N$ data points.
The log-likelihood for the distribution Eq.\~(\ref{Eq:PLnorm}) can be defined as \cite{Aban_2006, White_2008, Deluca_2013} 
\beq
\mathcal{L}(\mu) =  \sum_{i=1}^N \ln P(n_i)
= N \ln \left( \frac{1-\mu}{ n_{max}^{(1-\mu)}-1} \right) - \mu \sum_{i=1}^N \ln n_i .
\label{Eq:LogLikeFun}
\eeq
Then, an estimator for the cut-off value is obtained by $\hat{n}_{max} = \max (n_i)$ \cite{Aban_2006} and
an estimator $\hat{\mu}$
for the critical exponent is obtained by maximizing $\mathcal{L}(\mu)$, yielding a standard error \cite{Deluca_2013} 
\beq
\sigma = \frac{1}{\sqrt N}  \left[
\frac{1}{(\hat{\mu} -1)^2} -
\frac{\hat{n}_{max}^{(\hat{\mu} -1)}  \ln^2 \hat{n}_{max} }
{\left(1 - \hat{n}_{max} ^{( \hat{\mu} -1)} \right)^2 }
\right] ^{-\frac{1}{2}} .
\label{Eq:LogStd}
\eeq 
In these expressions, the 'hat' means that we refer to an estimated value.
The maximization of $\mathcal{L}(\mu)$ must be computed numerically (here we use Brent's method
from the Julia package Optim.jl \cite{Mogensen_2018}).
In the limit of an unbounded power-law distribution $n_{max}\to \infty$, the maximization can be calculated analytically, yielding
\cite{Newman_2005}
\begin{equation}
    \hat{\mu} = 1 + N \left( \sum_i \ln n_i   \right)^{-1} .
    \label{Eq:Loglike}
\end{equation}
%



To estimate the distribution $P(n)$ of case numbers that vary over many orders of magnitude, we
used a histogram with logarithmic binning. That is,  
we placed a discrete number of bins, $k$, at positions of integer powers of two $n_k = 2^k$ (i.e., 1,2,4,8,16, etc) and for each bin
counted the number $H_k$ of countries (or US counties)  that under the day of investigation
reported a number $n$ of cases that was falling into this bin ($n_{k} \leq n < n_{k+1}$).
To obtain the probability distribution, the resulting histogram counts were divided by the varying bin sizes 
$P(n_k) = c H_k / (n_{k+1} - n_{k})$ and the normalization constant $c$ fixed so that 
$\sum_k P(n_k) = 1$.
For visualization we plotted the distribution on double-logarithmic axes, 
excluding bins without entries. 

To confirm the robustness of the histogram estimation, we also used an alternative algorithm, where we
first computed the histogram of log-transformed 
case numbers $\nu = \log(n)$ using equally-spaced bins, 
which, after normalization, yielded the distribution $\tilde{P}(\nu)$.
Next, we  
used the back-transform $P(n) = \tilde{P}(\exp(\nu)) / n$ to obtain the probability distribution $P(n)$ of
non-logarithmic case numbers.
This procedure also yields a distribution with bins that are equally spaced on a logarithmic scale and
and the resulting distributions, shown in the Appendix Fig.~\ref{fig:distributions_all_a},
are very similar to that from the logarithmic binning method described above.
We have checked that the resulting distribution is largely independent to the choice and number of used histogram bins and other 
numerical parameters.

We also computed the cumulative fraction $C(n) =\sum_{m=n+1}^N  P(m)$ of countries with 
case number $m > n$.
This was obtained by taking a rank-plot of case numbers and inverting axes, 
i.e., sorting the array of case numbers in descending order and
plotting for each country the rank as a function of the sorted case number on double-logarithmic axes \cite{Newman_2005}.

\subsection*{The metapopulation model}

To describe the spatio-temporal evolution
of epidemic prevalence during the course of a pandemic
we developed a conceptual dual-scale meta-population model. 
The large-scale model component allows to simulate the spread of the virus in a network of $N$ interconnected countries. 
The state of a country is given as a Boolean value, being either invaded by the virus or non-invaded.
The model starts with a single invaded country. 
The geographic spread runs in discrete time, 
each step corresponding to one day of the time-continuous small-scale model. 
In each time step, every non-invaded country becomes infected by an invaded country with probability $p$. 
Thus, if at time $t$ a number of $m$ countries have already been invaded, the probability for a non-invaded 
country to receive the virus in this time step equals $1-(1-p)^m$.
As soon as a country has been invaded by the virus in this process, the small-scale model for this country is initiated.
Parameter values were taken as follows:
number of countries $N=200$ and
invasion probability $p=6\cdot 10^{-4}$. 

The small-scale model is time-continuous and deterministically
describes the epidemic dynamics within a country.
The model determines the time course of susceptible $S$, infected $I$, recovered $R$, and dead $D$ 
from a standard SIR-model \cite{Kermack_1927, Keeling_Rohani_2008}

\beq
\dot{S}  =  - \beta  \frac{S}{N_{pop}} I ,\
\dot{I} =  \beta \frac{S}{N_{pop}} I  - \gamma I, \
\dot{R}  =   (1-m) \gamma I, \
\dot{D}  =   m \gamma I.
\label{Eq:SIR}
\eeq

Here, $N_{pop}$ is the constant population size in the country, $\beta$ is the contact rate, $1/\gamma$ the infectious period, 
and $m$ the case fatality rate.
The total number of cases is determined as $C=I+R+D$. 
In the small-scale model countries are simulated independently from each other
and are only coupled by the unique invasion event for each country, 
which starts the epidemic growth in that country with 
initial values $S(0)=5\cdot 10^7$, $I(0)=1$ and $R(0)=D(0)=0$.
All infection state variables in a country are
zero before invasion by the virus, $I=R=D=0$.
The resulting well-known SIR-dynamics in a single country is shown in the Appendix Fig.~\ref{fig:SIRD}.
With the chosen parameterization, it takes roughly 80 days until the epidemic peak is reached. 
After this time, the assumption of an exponential increase, Eq.~(\ref{Eq:Exp}), breaks down.
Parameter values were taken as follows:
country population size $N_{pop}=5\cdot 10^7$,
case fatality rate $m=0.01$, 
infectious period $1/\gamma = 6 d$, and
contact rate $\beta=0.4 d^{-1}$.
This yields a growth rate $r=\beta - \gamma=0.23 d^{-1}$, corresponding to a doubling time of 
$T_{1,2}=\log(2)/r= 3 d$ 
and a  basic reproduction number $R_0 = \beta / \gamma = 2.4$.


\medskip

\subsection*{Acknowledgments}
I would like to thank Christoph Feenders, Thilo Gross, Alastair Jamieson-Lane, Cora Kohlmeier, James McLaren, 
and  Alexey Ryabov for helpful comments regarding the manuscript.\\

\subsection*{Software and data availability}
Code for data analysis and numerical simulations was written in Julia  \cite{Bezanson_2017}.
The differential equations were solved with the package
DifferentialEquations.jl \cite{Rackauckas_2017}, the maximization of the log-likelihood function with
package Optim.jl \cite{Mogensen_2018}.
The source code of the used algorithms will be made publicly available at
\url{https://github.com/berndblasius/Covid19}.
The used data are available at
\url{https://github.com/CSSEGISandData}.




\begin{figure}[t]
\center \includegraphics[width=0.8\textwidth]{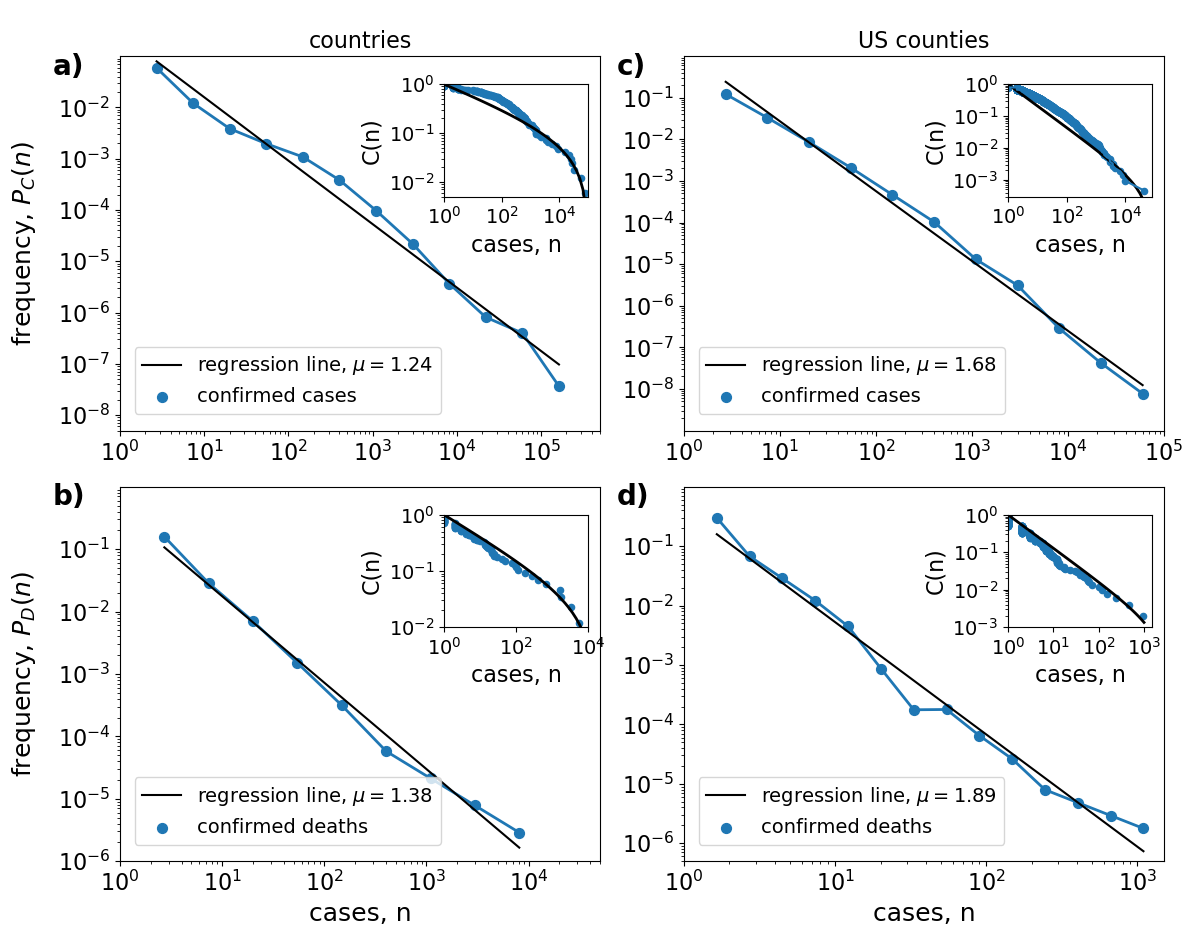}
\caption{
Alternative calculation of the distribution of COVID-19 case numbers in countries worldwide and
counties in the US.
Same as Fig.~\ref{fig:distributions_all}, but for the alternative histogram algorithm,
using a histogram of log-transformed case numbers (see Methods).
}
\label{fig:distributions_all_a}
\end{figure}

\begin{figure}[t]
\center \includegraphics[width=0.8\textwidth]{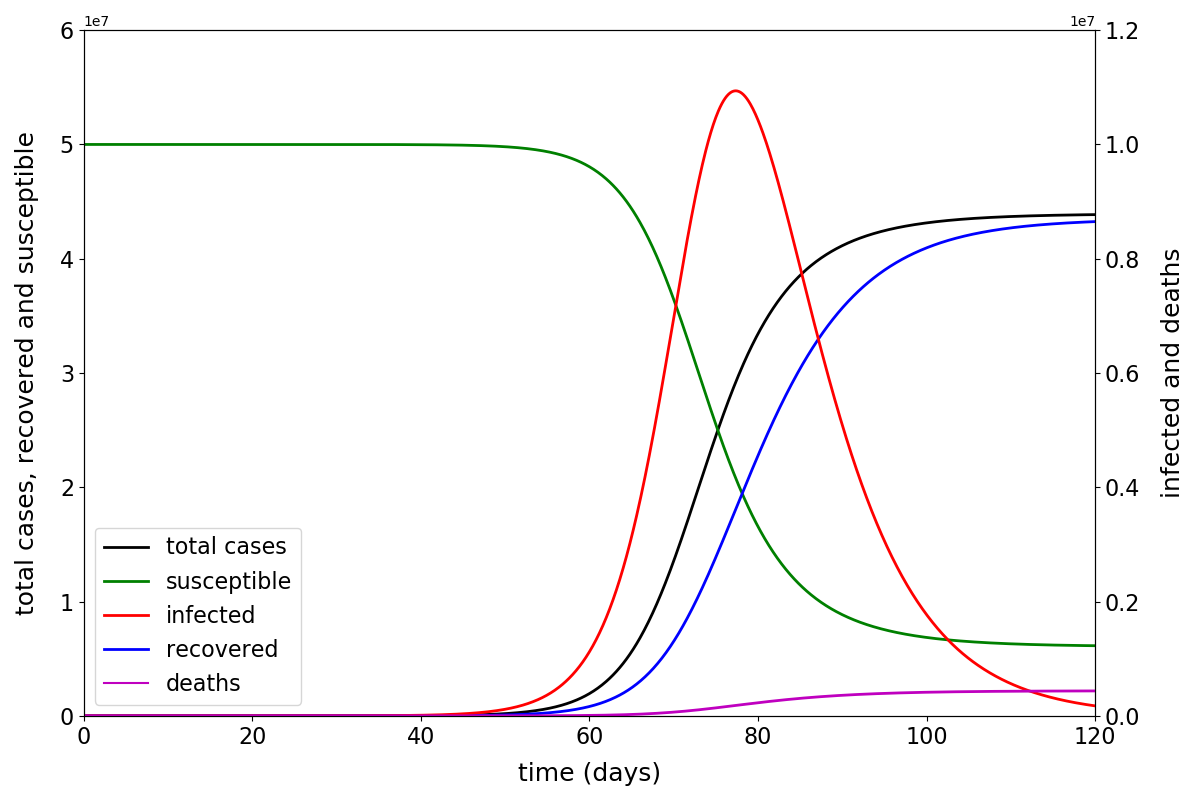}
\caption{
Simulation of the SIR-model (\ref{Eq:SIR}) within a country.
The plot shows the numerically obtained values of the total number of cases $C$ (black), 
the number of susceptible $S$ (green), and
the number of recovered $R$ (blue) on the left axis, as well as the number of infected $I$ (red) and deaths $D$ (magenta)
and the right axis as a function of time.
For the used initial values of $S(0)=5\cdot 10^7$, $I(0)=1$ and $R(0)=D(0)=0$, the epidemic peak is reached after 77 days.
See methods for parameter values.
}
\label{fig:SIRD}
\end{figure}

\begin{figure}[ht]
\center \includegraphics[width=0.8\textwidth]{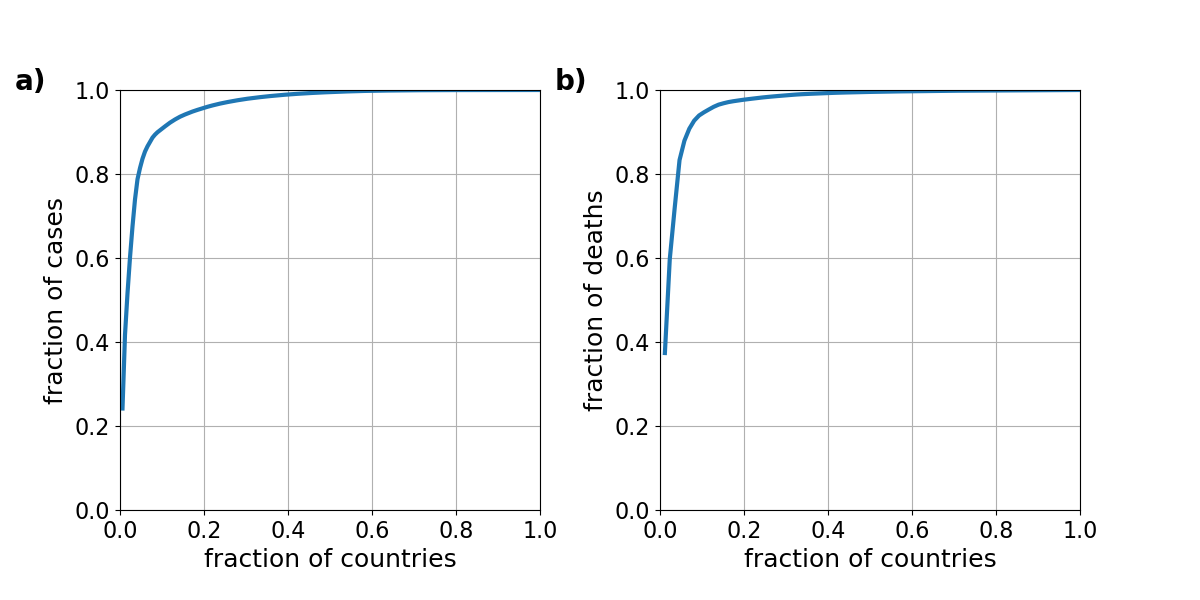}
\caption{
Lorenz curves, depicting the inequality in the distribution of confirmed COVID-19 cases.
The plots show the fraction of the number of confirmed cases {\bf (a)} and of the number of confirmed
deaths {\bf (b)} as a function of the fraction of most affected countries on March 22, 2020 
(compare to Fig.~\ref{fig:distributions_all}).
This inequality corresponds to a Gini-coefficient of $G=0.92$ for the distribution of confirmed cases and
of $G=0.94$ for the number of confirmed deaths.
}
\label{fig:lorenz}
\end{figure}

\end{document}